\documentclass[preprint,showpacs,preprintnumbers,amsmath,amssymb,showkeys]
{revtex4}

\usepackage{graphicx}% Include figure files
\usepackage{dcolumn}% Align table columns on decimal point
\usepackage{bm}% bold math

\begin{document}

\preprint{TIFR/TH/03-17}

\title{Strange Stars with a Density-Dependent Bag Parameter}
\author{Nirvikar Prasad}
 \email{nirvikar@mailhost.tifr.res.in}
\author{R.~S. Bhalerao}
 \email{bhalerao@theory.tifr.res.in}
\affiliation{Tata Institute of Fundamental Research, Homi Bhabha Road,
Colaba, Mumbai 400 005, India}

%\date{\today}% It is always \today, today,
             %  but any date may be explicitly specified
\date{January 23, 2004}% It is always \today, today,

\begin{abstract}
We have studied strange quark stars in the framework of the MIT bag
model, allowing the bag parameter $B$ to depend on the density of the
medium. We have also studied the effect of Cooper pairing among
quarks, on the stellar structure. Comparison of these two effects
shows that the former is generally more significant. We studied the
resulting equation of state of the quark matter, stellar mass-radius
relation, mass-central-density relation, radius-central-density
relation, and the variation of the density as a function of the
distance from the centre of the star. We found that the
density-dependent $B$ allows stars with larger masses and radii, due
to stiffening of the equation of state. Interestingly, certain stellar
configurations are found to be possible only if $B$ depends on the
density. We have also studied the effect of variation of the
superconducting gap parameter on our results.
\end{abstract}

% PACS, the Physics and Astronomy Classification Scheme.
%Use showkeys class option if keyword display desired
\maketitle

\section{\label{sec:level1}Introduction}

The conjecture that the true ground state of quantum chromodynamics
(QCD) may be the strange quark matter (SQM) was first explicitly
stated by Witten \cite{Witten:1984rs}, though various forms of quark
matter had been considered earlier
\cite{Bodmer:1971we,Chin:1979yb}. The key idea is that the energy per
baryon of SQM could be less than that of even $^{56}$Fe, the most
stable nucleus, making SQM comparatively more stable. Farhi and Jaffe
\cite{Farhi:1984qu} studied SQM in the framework of the MIT bag model
\cite{Chodos:1974je}, for various values of the strange quark mass
($m_s$) and the bag constant ($B$). They found a ``window of
stability" in the $m_s$-$B$ plane inside which SQM was stable. The
astrophysical implication of the stability of SQM, namely the possible
existence of strange (quark) stars, was studied by 
Haensel {\it et al.} \cite{Haensel:qb} and
Alcock {\it et al.} \cite{Alcock:1986hz}. They
found that very compact strange stars are possible, which are much
smaller than the normal neutron stars, the reason being that the
strange stars are self-bound and generically different from the
gravitationally bound stars. Of course, the actual existence of
strange stars can only be confirmed by observational studies of
pulsars and pulsar-like compact stars, including X-ray pulsars,
X-ray bursters, 
anomalous X-ray pulsars, soft $\gamma$-ray repeaters and
isolated neutron stars.

The SQM hypothesis has been around for almost two decades. Recent
theoretical progress in our understanding of the cold and dense QCD
matter has renewed the interest in this hypothesis: At low
temperatures and sufficiently high baryon number densities, it is
thought that quarks near the Fermi surface form Cooper pairs and new
condensates develop. This is expected to lead to a
colour-superconducting phase in the QCD phase diagram
\cite{Alford:1997zt,Rapp:1997zu}. Among the various
colour-superconducting phases, the colour-flavour-locked (CFL) phase
\cite{Rajagopal:2000wf,Alford:1999pa,Schafer:1999pb} is the one in
which quarks of all three colours and all three flavours ($u$, $d$,
$s$) pair. Electrical neutrality of bulk CFL matter is ensured due
to the fact that this phase favours equal number densities of $u$, $d$
and $s$ quarks, despite their unequal masses
\cite{Rajagopal:2000ff,Steiner:2002gx}. This also means that electrons
are absent in the bulk CFL matter. Strange stars made of CFL matter
seem to be favoured over those made of another colour-superconducting
phase, namely the 2SC phase \cite{Alford:2002kj}. These theoretical
developments have important implications for the existence of quark
stars and neutron stars with quark cores.

Other reasons for the revival of interest in this area are the recent
claims of observation of both mass and radius of compact stars
\cite{Pons:2001px,Drake:2002bj}. Although these claims are somewhat
controversial \cite{Walter:2002uq}, they have further stimulated
theoretical activity in this area.

The stability of SQM in the light of the above theoretical
developments, and the structure of strange stars have recently been
investigated by Lugones and Horvath
\cite{Lugones:2002va,Lugones:2002zd}. While constructing the equation
of state (EOS) of the CFL matter, they introduced the vacuum energy by
means of the phenomenological bag constant $B$. However, in view of
the high densities encountered in neutron stars and quark stars, the
question whether the bag constant should be held fixed at its
free-space value or whether its variation with the density of the
medium should be taken into account becomes very relevant. In this
work, we investigate the effect of a density-dependent $B$, on the EOS
and the resulting structure of the CFL strange stars.

The paper is organized as follows. In Sec. II we discuss various
models of $B(\rho)$ where $\rho$ is the baryon number density of the
medium. In Sec. III we consider the CFL matter and its EOS. Our
numerical results for the resulting stellar structure are given in
Sec. IV. In Sec. V we present our conclusions.

%-------------------------------------------------%
\section{Density-Dependent Bag Parameter}
%-------------------------------------------------%

Of the many models of the nucleon that have been constructed so far,
one of the most useful is the MIT Bag Model \cite{Chodos:1974je}: The
nucleon is assumed to be a ``bubble'' of perturbative vacuum immersed
in the non-perturbative or true vacuum, and quarks are confined to the
bubble by means of a net inward pressure $B$ exerted on it by the
surrounding vacuum. The bag model has also been used to study the
thermodynamics of the deconfinement phase transition. As the
temperature rises above the deconfinement temperature, there is no
difference between the two vacua, and the net inward pressure $B$ must
vanish. In other words, $B$ must be viewed as a temperature-dependent
quantity \cite{Muller:1980kf}. The analogy between the bag constant
$B$ and the condensation energy in the Nambu--Jona-Lasinio (NJL) model
at finite temperature was pointed out in Ref. \cite{Li:es}. It was
shown that as the temperature rises, this condensation energy
decreases and goes to zero at the transition temperature. This
provides a model of the temperature-dependent $B$. Density dependence
of $B$ can be motivated in an analogous manner: Since the
deconfinement and the concomitant vanishing of $B$ can also be brought
about by raising the baryon number density $\rho$, $B$ should be
treated as a density-dependent quantity \cite{Reinhardt:tv}.

There have been a number of attempts in the literature to evaluate the
density dependence of $B$. The results, however, are model dependent,
and no consensus seems to have emerged. In this work, we made use of
the results of the following three studies published recently:
(a) Liu {\it et al.} \cite{Liu:2001em} extended the global colour
symmetry model \cite{Roberts:1987xc} to finite quark number chemical
potentials ${\mu}$, in order to derive the variation of $B$ with
$\mu$ and $\rho$.
(b) Burgio {\it et al.} \cite{Burgio:2001mk} used the CERN SPS data on
heavy-ion collisions to justify and determine the density dependence
of $B$.
(c) Aguirre \cite{Aguirre:2002ws,Aguirre:2003pc} 
used the NJL model to study the
modification of the QCD vacuum with increasing baryonic density and to
extract the medium dependence of $B$.

(a) Since Liu {\it et al.} have not presented their $B(\rho)$
or $B(\mu)$ in a parametric form, we have fitted their results with
a suitable analytic expression with two parameters $a_1$ and $a_2$:
\begin{equation}
B(\rho)/B(0) = \exp[-(a_1 x^2 + a_2 x)],
\end{equation}
where $x \equiv \rho/\rho_0$ is the normalized baryon number density,
$\rho_0$ is the baryon number density of the ordinary nuclear matter,
$a_1=0.0125657$, and $a_2=0.29522$. They took $B(0)=114$ MeV/fm$^3$
=(172 MeV)$^4$.
(b) Burgio {\it et al.} have presented their $B(\rho)$ in a parametric
form:
\begin{equation}
B(\rho) = B_{as} + (B_o-B_{as})\exp(-\beta x^2),
\end{equation}
where $B_{as}=38$ MeV/fm$^3$, $B_0 = 200$ MeV/fm$^3$ =
(198 MeV)$^4$, $\beta = 0.14$,
and $x$ is as defined above.
(c) Aguirre \cite{Aguirre:2003pc} has calculated $B(\rho)$ for a
symmetric $uds$ quark matter relevant for a CFL quark star. 
His $B(\rho)$, in MeV/fm$^3$, is given by
\begin{eqnarray}
B(\rho) &=&
a + b_1 x + b_2 x^2 + b_3 x^3 + b_4 x^4 + b_5 x^5,  ~~~~ x \leq 9
\nonumber \\
&=& \beta \exp(-\alpha (x-9)),  ~~~~ x > 9
\end{eqnarray}
where $x$ is as defined earlier. Further, 
$a=291.59096,
~b_1=-142.25581,
~b_2=39.29997,
~b_3=-6.04592,
~b_4=0.46817,
~b_5=-0.01421,
~\alpha=0.253470705$, and $\beta=19.68764$.
Thus in this case $B(0)=291.59096$ MeV/fm$^3$ = (217.6 MeV)$^4$. 

We display, in Fig. 1, $B(\rho)$ vs $\rho$ corresponding to Eqs. (1)-(3).
These three models provide $B(\rho)$ over a sufficiently wide range of
densities expected to be seen inside quark stars. Secondly,
the above three values of $B(0)$ span more or less the whole range
of values of the bag constant found in the literature. Finally, as is
clear from Eqs. (1)-(3), they yield qualitatively different shapes for
$B(\rho)$: 
(1) and (3) have negative slopes at $x=0$, while
(2) has a vanishing slope; (1) and (3) vanish as $x \rightarrow
\infty$, while (2) tends to a positive constant. It will be
interesting to investigate what effects these different shapes 
and magnitudes of $B(\rho)$ have on
the EOS and the stellar structure. We shall use these three models in
subsequent sections; all our numerical results are labeled accordingly
as (a), (b), or (c).

%-------------------------------------------------%
\section{Thermodynamics of the CFL phase}
%-------------------------------------------------%

The free energy density $\Omega_{CFL}$ of the CFL phase, at
temperature $T=0$, is obtained as follows \cite{Rajagopal:2000ff}.
We first consider the free
energy density $\Omega_{free}$ of an unpaired quark matter in which
all quarks that are ``going to pair'' have a common Fermi momentum
$\nu$: the CFL pairing is most effective if the Fermi momenta of the
$u$, $d$, and $s$ flavours are the same.
Although the nonzero strange quark mass tends to favour unequal number
densities for the three flavours, the CFL pairing forces the three
flavours to have the same Fermi momentum and hence the same number
density, so long as $m_s$ is not too large. If $\Delta$ is the
colour-superconductivity gap parameter and $\mu$ is the quark number
chemical potential, then the binding energy of the $qq$ Cooper pairs
can be included (to leading order in $\Delta/\mu$) by subtracting
$3\Delta^2 \mu^2 /\pi^2$ from $\Omega_{free}$. The vacuum energy
density in the presence of a finite chemical potential $\mu$, is
included by adding $B(\mu)$. This gives
\begin{eqnarray}
\Omega_{CFL} &=& \Omega_{free} - \frac{3}{\pi^2} \Delta^2 \mu^2 + 
B(\mu) \nonumber \\
&=& \frac{6}{\pi^2} \int_{0}^{\nu} (p - \mu) p^2 dp
+ \frac{3}{\pi^2} \int_{0}^{\nu} [(p^2 + m_s^2)^{1/2} - \mu] p^2 dp
- \frac{3}{\pi^2} \Delta^2 \mu^2 + B(\mu),
\label{om1}
\end{eqnarray}
see \cite{Alford:2002rj,Alford:2001zr} which, however, treat $B$ as a
constant. The baryon number density $\rho=-\partial \Omega_{CFL}/(3
\partial \mu)$ and particle number densities for individual flavours
are given by
\begin{equation}
\rho = n_u = n_d = n_s = \frac{(\nu^3 + 2 \Delta^2 \mu)}{\pi^2}
-\frac{1}{3}B'(\mu),
\label{nb}
\end{equation}
where $B'(\mu) \equiv d B / d \mu.$
The common Fermi momentum $\nu$ is chosen to minimize the free energy
density with respect to a variation in $\nu$. This gives
\begin{equation}
\nu = 2 \mu - \left( \mu^2 + \frac{m_s^2}{3} \right)^{1/2} \label{nu}.
\end{equation}
Energy density of the CFL matter is given by
\begin{equation}
\varepsilon = \sum_i \mu_i  n_i  + \Omega_{CFL} = 3 \mu \rho - P,
\label{E}
\end{equation}
since $\mu = (\mu_u + \mu_d + \mu_s)/3$ and the pressure $P$ is
given by $P=-\Omega_{CFL}$.

Now we obtain the equation of state of the CFL matter, in the standard
form. Integrations in Eq. (4) can be performed in a straightforward
manner and we get
\begin{equation}
\Omega_{free}=\frac{3 \nu^4}{2 \pi^2}-\frac{3 \mu \nu^3}{\pi^2}
+\frac{3}{\pi^2}\left[\frac{\nu}{4}(m_s^2+\nu^2)^{3/2}
-\frac{m_s^2 \nu}{8}(m_s^2+\nu^2)^{1/2}-\frac{m_s^4}{8}\sinh^{-1}
\frac{\nu}{m_s} \right].
\end{equation}
We are interested in the region where $m_s$ is small compared to the
chemical potential $\mu$. So it is sufficient to keep terms up to
order $m_s^4$ in $\Omega_{free}$ \cite{Alford:2002kj}. To that end,
first note that the common Fermi momentum $\nu$ (Eq. (6)) is given
by
\begin{equation}
\nu  =  \mu - \frac{m_s^2}{6 \mu} + \frac{m_s^4}{72 \mu^3},
\end{equation}
where we have kept terms up to order $m_s^4$. This yields
\begin{equation}
\Omega_{CFL} = \frac{- 3 \mu^4}{4 \pi^2}
+ \frac{3 m_s^2 \mu^2}{4 \pi^2}
- \frac{1 -12 \log(m_s /2 \mu)}{32 \pi^2} m_s^4
- \frac{3}{\pi^2} \Delta^2 \mu^2 + B(\mu).
\end{equation}
Substitution of $\nu$ given by Eq. (9), in Eqs. (5) and (7) yields
\begin{equation}
\rho =  \frac{\mu^3}{\pi^2 } - \frac{m_s^2 \mu}{2 \pi^2 }
+ \frac{m_s^4 }{8 \pi^2 \mu }
+ \frac{2}{\pi^2} \Delta^2 \mu -\frac{1}{3}B'(\mu),
\end{equation}
\begin{equation}
\varepsilon = \frac{9 \mu^4}{4 \pi^2} - 
\frac{3 m_s^2 \mu^2}{4 \pi^2 } + \frac{11 m_s^4}{32 \pi^2 }
+ \frac{3 \log(m_s /2 \mu)}{8 \pi^2}  m_s^4
+ \frac{3}{\pi^2} \Delta^2 \mu^2 + B(\mu)-\mu B'(\mu).
\label{eaprox}
\end{equation}
The pressure is given by
\begin{equation}
P =
\frac{3 \mu^4}{4 \pi^2 } - \frac{3 m_s^2 \mu^2}{4 \pi^2}
+ \frac{1 -12 \log(m_s /2 \mu)}{32 \pi^2} m_s^4
+ \frac{3}{\pi^2} \Delta^2 \mu^2 - B(\mu).
\label{paprox}
\end{equation}

Equations (\ref{eaprox}) and (\ref{paprox}) give $\varepsilon$ and $P$
in terms of $\mu$ as a parameter. The EOS in the standard form can be
obtained from these two equations and it reads
\begin{equation}
\varepsilon = 3P + 4B(\mu)
- \frac{6 \Delta^2 \mu^2}{\pi^2} + \frac{3 m_s^2 \mu^2}{2 \pi^2}
+ \frac{1 +6 \log(m_s /2 \mu)}{4 \pi^2} m_s^4-\mu B'(\mu).
\label{eos}
\end{equation}
If only the first two terms on the rhs of Eq. (\ref{eos}) are kept
then it reduces to the ideal gas EOS in the bag model. The occurrence
of the pairing term stiffens the EOS because we get a higher pressure
at the same energy density. A nonzero value of $m_s$, on the other
hand, tends to reduce the pressure leading to a softer EOS.

If Eq. (14) is to be used to study the stellar structure, one needs to
know $B(\mu)$ and its derivative $B'(\mu)$. In Sec. II, we have
described three models of $B(\rho)$. Consider first the model of Liu
{\it et al.} \cite{Liu:2001em}, denoted by (a) in Sec. II. Their
formalism in fact first yields $B(\mu)$, not $B(\rho)$. They, however,
do not present their numerical results for $B(\mu)$. Instead, they use
the relation $\rho=2 \mu^3/(3 \pi^2)$ to obtain $B(\rho)$ and present
numerical results for $B(\rho)$. In order to extract their $B(\mu)$,
we have to use exactly the same relation between $\rho$ and $\mu$.
This is how we obtained $B(\mu)$ in the case of the model (a). Next,
consider the models (b) and (c) of Sec. II. To obtain $B(\mu)$ from
$B(\rho)$ in these cases, we proceed as follows. Consider Eq. (11).
It can be rewritten as
\begin{equation}
\rho=f(\mu)-\frac{1}{3}\frac{dB}{d\mu},
\end{equation}
where $f(\mu)$ denotes the sum of the first four terms on the rhs of
Eq. (11). In the case of model (b), Eq. (2) can be inverted to get
$\rho(B)$ which can be substituted in Eq. (15). This gives a
first-order differential equation between $B$ and $\mu$, which
can be solved numerically. In the case of model (c), it is 
easier to eliminate $B$ between Eqs. (3) and (15), and the
resulting differential equation looks like
\begin{equation}
\frac{dx}{d\mu}=3 [f(\mu) - \rho_0 x ] / (dB / dx).
\end{equation}
In both the cases, the initial condition necessary to solve the
differential equation is obtained from the fact that for sufficiently
large $\mu$, $dB/d\mu$ is nearly zero. We have checked the solutions
of the above differential equations by substituting them in the
original equations (Eqs. (15) and (2) in the case of model (b), and
Eqs. (15) and (3) in the case of model (c)).

Before we display our results for the EOS, Eq. (14), it is useful to
see how the quantities $B$, $B'$, $\varepsilon$ in Eq. (12), and $P$
in Eq. (13) vary with $\mu$. This is shown in Fig. 2. Consider first
$B(\mu)$. Its broad features are in agreement with those of $B(\rho)$
seen in Fig. 1. Also note that in spite of the large differences among
the values of $B(\rho=0)$ corresponding to the three models, the
curves for $B(\mu)$ in the region relevant to quark stars ($\mu \sim$
a few hundred MeV) do not differ drastically from each other.
Consider next $B'(\mu)$. It is either negative or zero depending on
the value of $\mu$. We have plotted $|\mu B'(\mu)|$ since it occurs in
Eq. (12). This quantity obviously tends to vanish at very small and
very large values of $\mu$, and it displays a gentle peak at an
intermediate value of $\mu$. $\varepsilon(\mu)$ contains $|\mu
B'(\mu)|$, whereas $P(\mu)$ does not. However, because of the
relatively small magnitude of $|\mu B'(\mu)|$, 
the curves for $\varepsilon(\mu)$ and
$P(\mu)$ have similar shapes. Interestingly, the magnitudes of
$\varepsilon(\mu)$ and $P(\mu)$ are seen to be
insensitive to the details
of the three models of $B(\rho)$. Finally, note that $P(\mu)$ is
positive only when $\mu$ exceeds a certain threshold value.

Figure 3 displays the EOS of the CFL matter and unpaired quark matter,
in the various models considered above. It is easy to understand the
shapes of these curves by looking at $P(\mu)$ and $\varepsilon(\mu)$
in Fig. 2, and noting that Fig. 3 is obtained essentially by
eliminating $\mu$. Some general trends emerge:
(1) If $B$ is treated as a constant independent of the density of the
medium then the EOS appears to be linear; see the dashed lines in
Fig. 3. On the other hand, if $B$ depends on $\mu$, the EOS is
slightly nonlinear.
(2) The pairing term is seen to make the EOS stiffer, so
does the density dependent $B$. However, the latter effect is seen to
be more significant. This will be corroborated further when we present
the stellar structure curves in the next section.

%\newpage

We now discuss the issue of the ``window of stability''. This is the
region in the $m_s$-$B$ plane, in which the SQM has energy per baryon
($\varepsilon / \rho$) less than 939 MeV, and which was first discussed
in Ref. \cite{Farhi:1984qu}. Since both $m_s$ and $B$ tend to increase
$\varepsilon / \rho$, this region obviously lies near small values of
$m_s$ and $B$, in the first quadrant of the $m_s$-$B$ plane. It is
bounded on the left by the vertical line 
$B=57$ MeV/fm$^3$=(145 MeV)$^4$, and
on the other side, by the $\varepsilon / \rho=939$ MeV contour. The
boundary on the left arises because if $B$ were smaller than 
57 MeV/fm$^3$, 
i.e., if the inward pressure holding the quarks together in the
nucleon were too small (see Sec. II), the nucleon and hence the nuclei
would have dissolved into $ud$ quark matter. Since that does not
happen, $B$ must not be smaller than 57 MeV/fm$^3$.

Recently, it was shown that the window of stability is enlarged
considerably if the SQM undergoes a phase transition to a CFL quark
matter \cite{Lugones:2002va}. The enlargement occurs because the
$\varepsilon / \rho=939$ MeV contour gets shifted to the right. Let us
understand this. To probe the absolute stability of the SQM, we need
to set the external pressure $P$ equal to zero. Then Eqs. (7) and (10)
yield, respectively,
\begin{equation}
\varepsilon / \rho = 3 \mu,
\end{equation}
and
\begin{equation}
\frac{- 3 \mu^4}{4 \pi^2}
+ \frac{3 m_s^2 \mu^2}{4 \pi^2}
- \frac{3}{\pi^2} \Delta^2 \mu^2 + B(\mu)= 0,
\end{equation}
where we have ignored the ${\cal O} (m_s^4)$ term in Eq. (10), for
simplicity. Ignoring the $\mu$-dependence of $B$ as in 
\cite{Lugones:2002va} and 
eliminating $\mu$ in the last two equations one gets
$\varepsilon / \rho$ as a function of $m_s$, $B$, and $\Delta$. It is
clear that a nonzero value of $\Delta$ shifts the $\varepsilon /
\rho=939$ MeV contour to the right thereby enlarging the window of
stability. This is essentially because of the opposite signs of the
$B$- and $\Delta$-terms in Eq. (18): The $\Delta$-term tends to
nullify the effect of a large $B$. The larger the $\Delta$, the larger
the value of $B$ one may have, while still maintaining $\varepsilon /
\rho < 939$ MeV. This is clearly borne out by the results of
\cite{Lugones:2002va}.

Let us now discuss how a $\mu$-dependent $B$ would affect the window
of stability. Earlier, when $\Delta$ was introduced, the functional
dependence of $\varepsilon / \rho$ on $m_s$ and $B$ was changed (see
Eqs. (17)-(18)), and hence the contour $\varepsilon / \rho=939$ MeV got
shifted. Now, whether $B$ is constant or $\mu$-dependent, the new
functional dependence remains unchanged. In other words, the
$\varepsilon / \rho=939$ MeV contour in the present work is the same as
in \cite{Lugones:2002va}.
We shall revisit the issue of the window of
stability in the next section, when we discuss Fig. 4.

\section{Structure of Strange Stars}

The equations of stellar structure, namely the
Tolman-Oppenheimer-Volkoff (TOV) equations
\cite{Tolman,Oppenheimer:1939ne}, give rise to a one-parameter family
of stars corresponding to a particular equation of state: By
specifying the central density as the parameter, one can numerically
integrate the TOV equations, starting at the centre of the star and
going radially outward until the pressure becomes zero which indicates
that the surface of the star has been reached. This
determines the radius $R$ and the mass $M$ of the star, for the
specified central density. By choosing successively larger values of
the central density, one can generate a sequence of stars of
increasing mass. The sequence ends when any further increase in the
central density leads to a star with a lower mass, which indicates
an unstable configuration. This end-of-sequence star is the maximum-mass 
star for the sequence. For a given
EOS, the mass-radius relationship is the most important property of
the family of stars. This is because it can be compared directly with
observational data in order to test and calibrate the theory. While
both the mass and the radius are not known so far for any pulsar, a
number of mass measurements exist and the data on the radius are
getting better.

We now present our numerical results on the structure of the strange
stars. We set the strange quark mass $m_s=150$ MeV and the gap
parameter $\Delta=0$ or 100 MeV (except in Figs. 8 and 9 where we vary
$\Delta$ between 0 and 150 MeV). We recall from
Ref. \cite{Lugones:2002va} that the stability of SQM requires that if
$m_s=150$ MeV and $\Delta=100$ MeV, then the bag constant $B$ should
be less than about 117 MeV/fm$^3$, and 
if $m_s=150$ MeV and $\Delta=0$, then
$B$ should be less than about 77 MeV/fm$^3$. 
Now consider the three models
of the density-dependent $B$, described in Sec. II. Since $B$ is a
decreasing function of the baryon number density $\rho$, the above
upper limits on the values of $B$ necessarily imply lower limits on
the values of $\rho$. In the numerical work, one has to ensure that
these limits are satisfied, so that the resulting star is always within
the window of stability.

We now discuss Figs. 4-9. In all these figures, dashed curves
correspond to the case when $B$ is taken as a constant independent of
the density of the medium, while the solid curves correspond to the
case when $B$ varies with $\rho$, and the panels (a), (b), (c) refer
to the three models of $B(\rho)$, described in Sec. II.

Figure 4 shows the mass-radius sequences. Consider first
Fig. 4(a). Model (a) has $B(0)=114$ MeV/fm$^3$ which is well within
the window of stability if $\Delta=100$ MeV (see above). This allows
us to draw the two curves labelled 100. {\it They show that a
density-dependent $B$ gives rise to larger and more massive stars.}
If $\Delta=0$, the above
value of $B(0)$ is outside the window. Hence there is no dashed curve
labelled 0. However, the solid curve labelled 0 is allowed: Recall
that the density cannot be arbitrarily small since the pressure needs
to be positive (Fig. 2). Indeed, for any point on this curve, the
density throughout the star is found to be high enough that $B(\rho)$
is everywhere less than 77 MeV/fm$^3$ (see above). {\it Interestingly,
this sequence of stars has become possible only because we considered
a density-dependent $B$.}
Now consider Fig. 4(b). Model (b) has
$B(0)=200$ MeV/fm$^3$ which is outside the window of stability for
both $\Delta=0$ and 100 MeV. Hence there are no dashed curves
here. However, the solid curves labelled 0 and 100, are allowed
because for any point on these curves, the density throughout the star
is found to be high enough that $B$ remains within the window of
stability. Note again that these sequences of stars have become
possible only because we considered a density-dependent $B$. Similar
remarks can be made about Fig. 4(c). {\it Results in Fig. 4 seem 
to be robust --- not very sensitive to the details of the three models
of $B(\rho)$ considered in this paper.}

The dot-dashed lines crossing the $M$-$R$ curves in Fig. 4 correspond
to the $M/M_\odot \simeq 0.15 ~R$ (in Km) constraint obtained by
Cottam {\it et al.} \cite{Cottam:2002cu} by using measurements of the
red shift of spectral lines in X-ray bursts from EXO0748-676. In
all Figs. 4(a-c) the crossing occurs at $M/M_\odot \simeq 1.3$-1.4.
{\it Thus the present
calculation has been able to construct strange stars which are
compatible with this constraint.}

It is a built-in feature of the TOV equations that the sequence of
stars ends, whatever may be the EOS. The typical shape of the sequence
shown in Fig. 4 ($M \propto R^3$ approximately) is generic to compact
stars that are self-bound. This is borne out by the fact that although
the EOS used here is quite different from that used in
\cite{Haensel:qb}, \cite{Alcock:1986hz} or \cite{Lugones:2002zd}, 
the shapes of the
curves in Fig. 4 are similar to those obtained in the earlier studies.

Density-dependent $B$ makes the EOS stiffer; see Fig. 3. A stiffer EOS
can resist the collapse of a star to a black hole, to a greater
extent. Hence stars with larger radii and masses result. What is true
for an individual star holds also for a sequence of stars, which as a
whole shifts toward larger mass and radius. This is evident in
Fig. 4(a).

Further, since the EOS is not strictly
linear, simple scaling where the
mass and radius scale as 1/$\sqrt{B}$ no longer holds
\cite{gle}. Actually when $\mu$-dependent $B$ is considered, the
meaning of scaling as such is lost because as one moves from the
surface of a star toward its centre, $B$ keeps decreasing. That is, no
fixed value can be assigned to it.

Finally, in Fig. 4(a), note that the difference between the two curves
labelled 100 is more significant than the difference between the two
solid curves. Similar observations can be made about Figs. 5-7. {\it
This underscores the importance of the modification of $B$ in a dense
medium vis-\`a-vis the phenomenon of Cooper pairing among quarks.}

We plot mass $M$ vs central energy density $\varepsilon_c$, in
Fig. 5. As $\varepsilon_c$ rises, so does the mass, until a plateau is
reached which leads to the end of the sequence. The slope
$dM/d{\varepsilon}_c$ is positive signifying stable configurations.
Since the solid curves correspond to a stiffer EOS (Fig. 3), for a
given value of the central energy density they represent more massive
stars. Figure 6 shows the corresponding variation of the radius $R$
with $\varepsilon_c$.

In Fig. 7 we show, for the maximum-mass star, the variation of the
density with the distance $r$ from the centre of the star. In the case
of the density-dependent $B$, the EOS is stiffer (Fig. 3) and the
central density is lower. In other words, the solid curve labelled 100
in Fig. 7(a) begins at a lower value of $\rho/\rho_0$ than the dashed
curve. In spite of this, the star in the former case is more massive
because it has a larger radius, and the large-$r$ regions contribute
to the mass in a dominant fashion.

We display the variation of the maximum mass with the gap parameter
$\Delta$, in Fig. 8. The increase in mass with $\Delta$ is expected
because as $\Delta$ increases the EOS becomes stiffer.
Recall that in model
(a), $B(0)=114$ MeV/fm$^3$. For this point to be inside the window of
stability, $\Delta$ has to be at least $\sim 100$ MeV. Hence in
Fig. 8(a) the dashed curve starts near $\Delta=100$ MeV. In Fig. 8(b)
there is no dashed curve because in model (b), $B(0)=200$ MeV/fm$^3$
which is not inside the window if $\Delta \le 150$ MeV. A similar
remark can be made about Fig. 8(c).
In Fig. 9 we plot the maximum radius vs $\Delta$. Recent data hint at
the possibility of very compact stars 
\cite{Pons:2001px,Drake:2002bj}. A smaller and denser
object in comparison to the neutron stars will be a strong candidate
for a strange star.

\section{Conclusions}

The fact that quarks normally exist in a confined state, does not
automatically rule out the strange quark matter hypothesis. This
hypothesis does not violate any physical principle and is also not
ruled out by any observed phenomenon. The real difficulty in verifying
the hypothesis lies in the fact that the underlying theory, namely the
QCD at quark chemical potential ($\mu$) of the order of several 100
MeV, is difficult to handle and does not yet allow a rigorous
calculation. Unfortunately, the phenomenological models which one
resorts to, give the energy per baryon of SQM very close to that of
$^{56}$Fe. In view of the uncertainties and the approximations
involved, it is difficult to make a definitive statement.

One way to make progress in this situation is to make the models more
realistic. Taking the density dependence of $B$ into account is a step
in that direction. 
\begin{enumerate}
\item 
We have shown that this has a significant effect on the equation of
state of SQM and hence on the resulting stellar structure curves.
Specifically, the density dependence of $B$ leads to a stiffer EOS
which gives rise to stars with larger masses and radii.
\item
Our results for the $M$-$R$ curves do not depend sensitively on the
details of the three models of the density dependence of $B$
considered in this paper.  
\item
Effect of the newly discovered phenomenon of colour superconductivity
on the structure of compact stars has been studied by several authors.
An important point which emerges from the present work is that the
results depend more sensitively on the density dependence of the
bag parameter than on the phenomenon of Cooper pairing of the quarks.
\item
Another observation that we made is with regard to the `window of
stability' of SQM. We showed that the density dependence of $B$ does
not affect the window.
\item
We have been able to construct stars that are compatible with the
constraint given in \cite{Cottam:2002cu}. Moreover, 
their masses lie in the astrophysically plausible
range.
\item
Most importantly, certain stellar sequences which are ruled out by the
requirement of the stability of SQM, if $B$ is treated as a constant,
are allowed if the density-dependence of $B$ is taken into
consideration.
\end{enumerate}

There are a number of ways in which this work can be extended.  
As mentioned in Sec. III, we considered the pairing interaction to the
leading order in $\Delta/\mu$. The same approximation was made also in
Refs. \cite{Alford:2002kj,Lugones:2002va,Lugones:2002zd,Alford:2002rj,
Alford:2001zr}. Since $\Delta/\mu$ can be as large as $\sim 1/3$, it
is clearly desirable to go beyond this approximation.
Secondly, the density dependence of $\Delta$ and $m_s$ can be included
to get a more complete picture. (In the present work, we have varied
$\Delta$ over a wide range and studied how it affects our results. It
appears that allowing $\Delta$ to depend upon the density of the
medium, may not change the results significantly. See, e.g., the two
solid curves in Fig. 7(a), where we show the variation of the density
as one goes from the center to the surface of the star. Each solid
curve is for a {\it fixed} value of $\Delta$, namely 0 or 100
MeV. Obviously, curves for intermediate values of $\Delta$ (say, 10,
20, ..., 90 MeV) would lie in between these two curves. Hence if
$\Delta$ were allowed to vary with the density or equivalently with
$r$, the resulting curve is not expected to be drastically different
from the two solid curves shown here.)
Thirdly, existence of additional phases (like 2SC, crystalline, etc.)
can be looked into. 
Finally, we did not include perturbative corrections to the equation
of state; these can be incorporated. 
Furthermore, a similar study where the bag parameter is dependent on
density can be made for hybrid stars.

We conclude by noting that the limitations of the present work,
pointed out in the preceding paragraph, and the simplicity of the
theoretical model employed, should be kept in mind while interpreting
the numerical results presented here.

\begin{acknowledgments}
N. Prasad acknowledges many helpful discussions with H.M. Antia and
A. Ray.
\end{acknowledgments}

\newpage
\begin{center}
Figure Captions
\end{center}

Fig. 1: Bag parameter $B$ vs baryon number density $\rho$
for the three models considered in this paper.

Fig. 2: Energy density $\varepsilon$, pressure $P$, bag parameter $B$,
and $|\mu ~dB/d \mu|$ vs quark number chemical potential $\mu$, for
the gap parameter $\Delta=100$ MeV. Here and in the rest of the
figures, the panels (a), (b), and (c) refer to the three models of the
density-dependent bag parameter $B(\rho)$, described in Sec. II. (For
the sake of clarity, $B \times 5$ rather than $B$ is plotted.)

Fig. 3: Pressure $P$ vs energy density $\varepsilon$, i.e., the
equation of state of the quark matter. In each panel, the two solid
curves correspond to a calculation with a density-dependent $B$, and
the two dashed curves correspond to a calculation with a constant
$B$. Curves are labelled by the values of $\Delta$ in MeV.

Fig. 4: Mass $M$ (in units of the solar mass $M_\odot$) vs radius $R$
of a strange star. Notation as in Figs. 2 and 3. The dot-dashed line
is the $M/M_\odot \simeq 0.15 ~R$ (in Km) constraint obtained in
Ref. \cite{Cottam:2002cu}.

Fig. 5: Mass $M$ (in units of the solar mass $M_\odot$) vs central
energy density $\varepsilon_c$ (in units of the energy density
$\varepsilon_0$ of the normal nuclear matter) of a strange
star. Notation as in Figs. 2 and 3.

Fig. 6: Radius $R$ vs central energy density $\varepsilon_c$ (in units
of the energy density $\varepsilon_0$ of the normal nuclear matter) of
a strange star. Notation as in Figs. 2 and 3.

Fig. 7: Baryon number density $\rho$ (in units of the baryon number
density $\rho_0$ of the normal nuclear matter) vs distance $r$, for
the maximum-mass strange star. Notation as in Figs. 2 and
3.

Fig. 8: The maximum mass $M$ of a strange star (in units
of the solar mass $M_\odot$) vs the gap parameter $\Delta$. Notation
as in Figs. 2 and 3.

Fig. 9: The maximum radius $R$ of a strange star vs the
gap parameter $\Delta$. Notation as in Figs. 2 and 3.

%\newpage

\begin{center}
\begin{figure}[ht!]
\centerline{\resizebox{14cm}{!} {\includegraphics{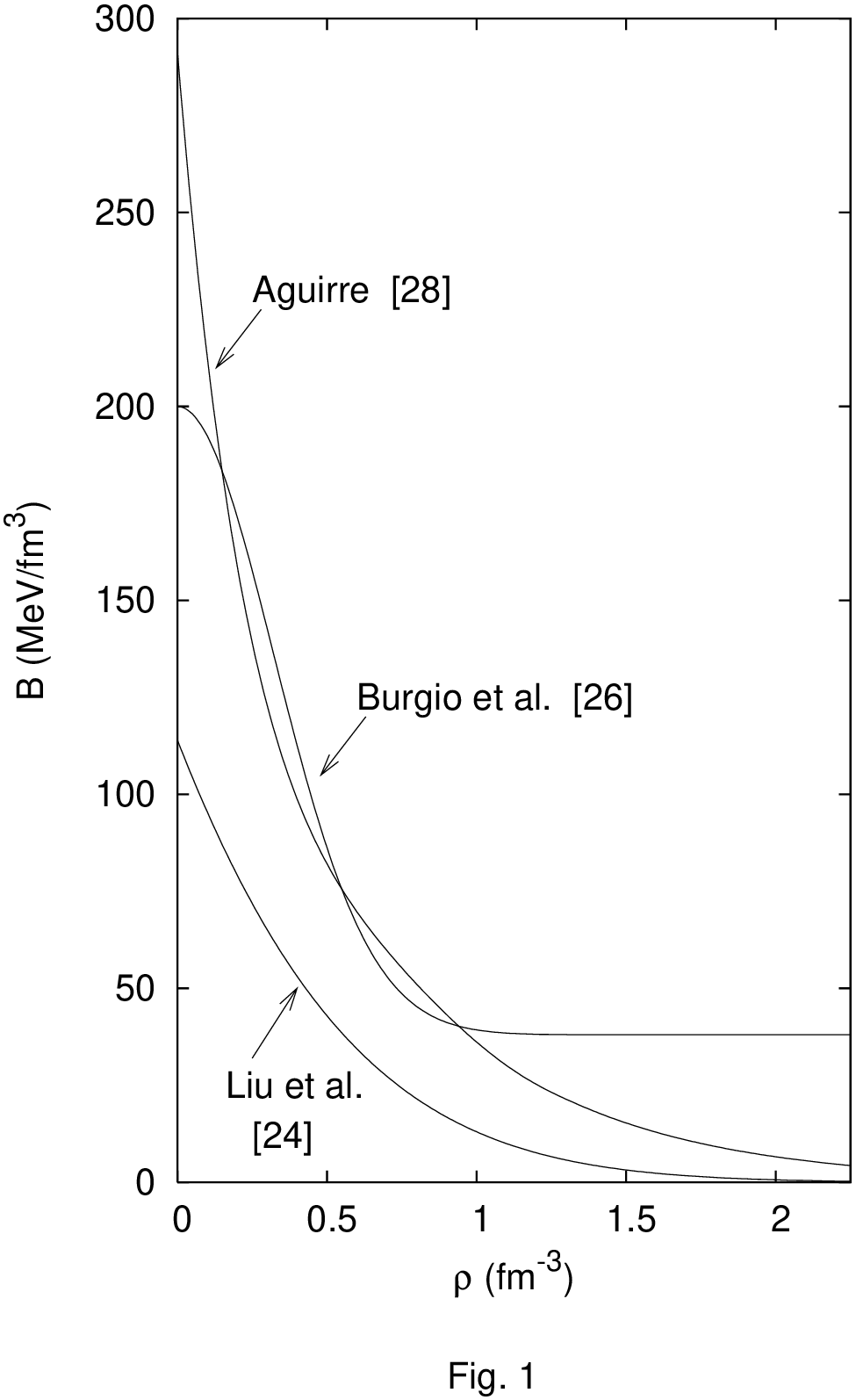}}}
%\caption{}
%\label{fig:...}
\end{figure}
\end{center}

\begin{center}
\begin{figure}[ht!]
\centerline{\resizebox{14cm}{!} {\includegraphics{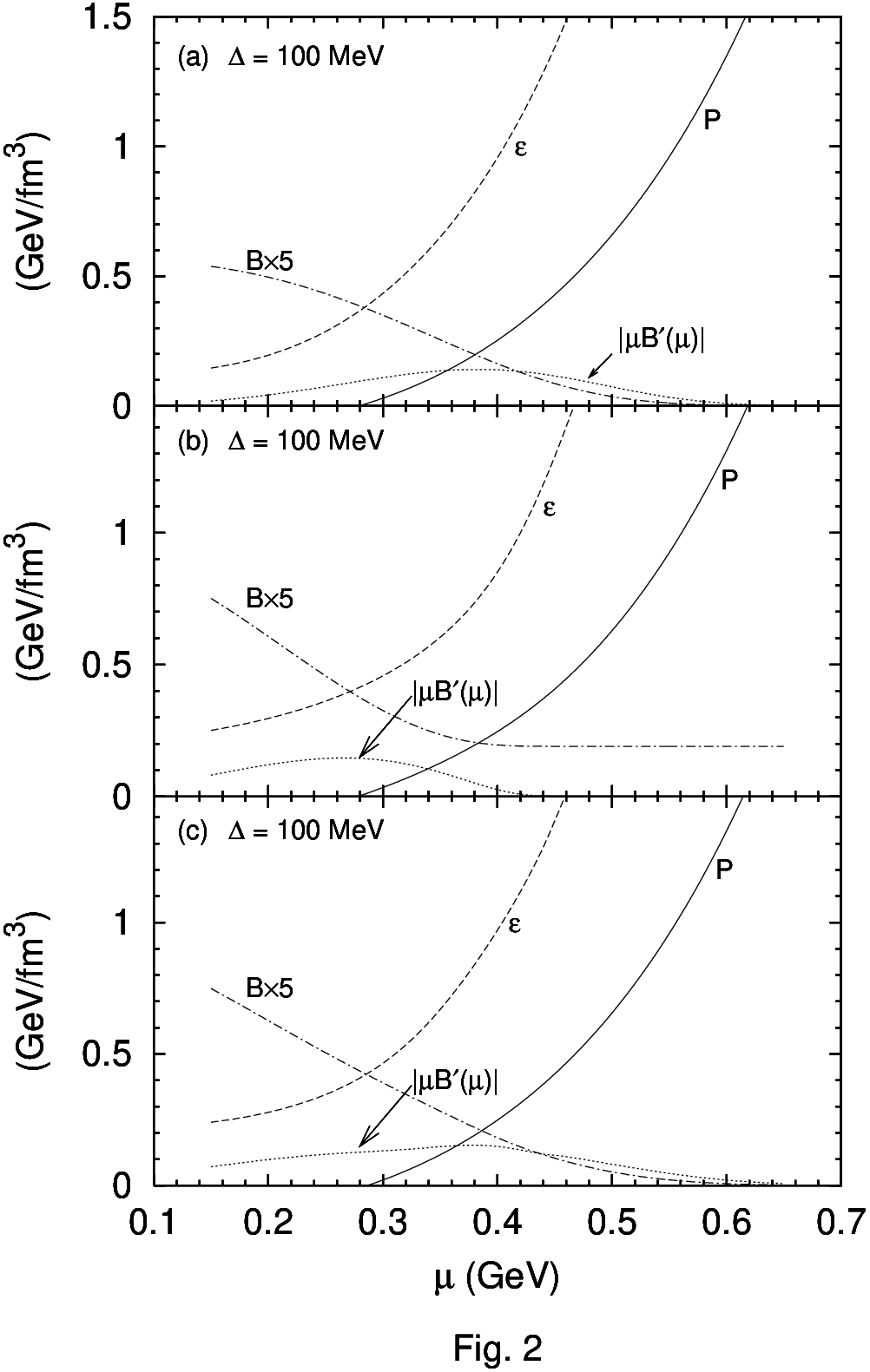}}}
%\caption{}
%\label{fig:...}
\end{figure}
\end{center}

%\newpage
\begin{center}
\begin{figure}[ht!]
\centerline{\resizebox{14cm}{!} {\includegraphics{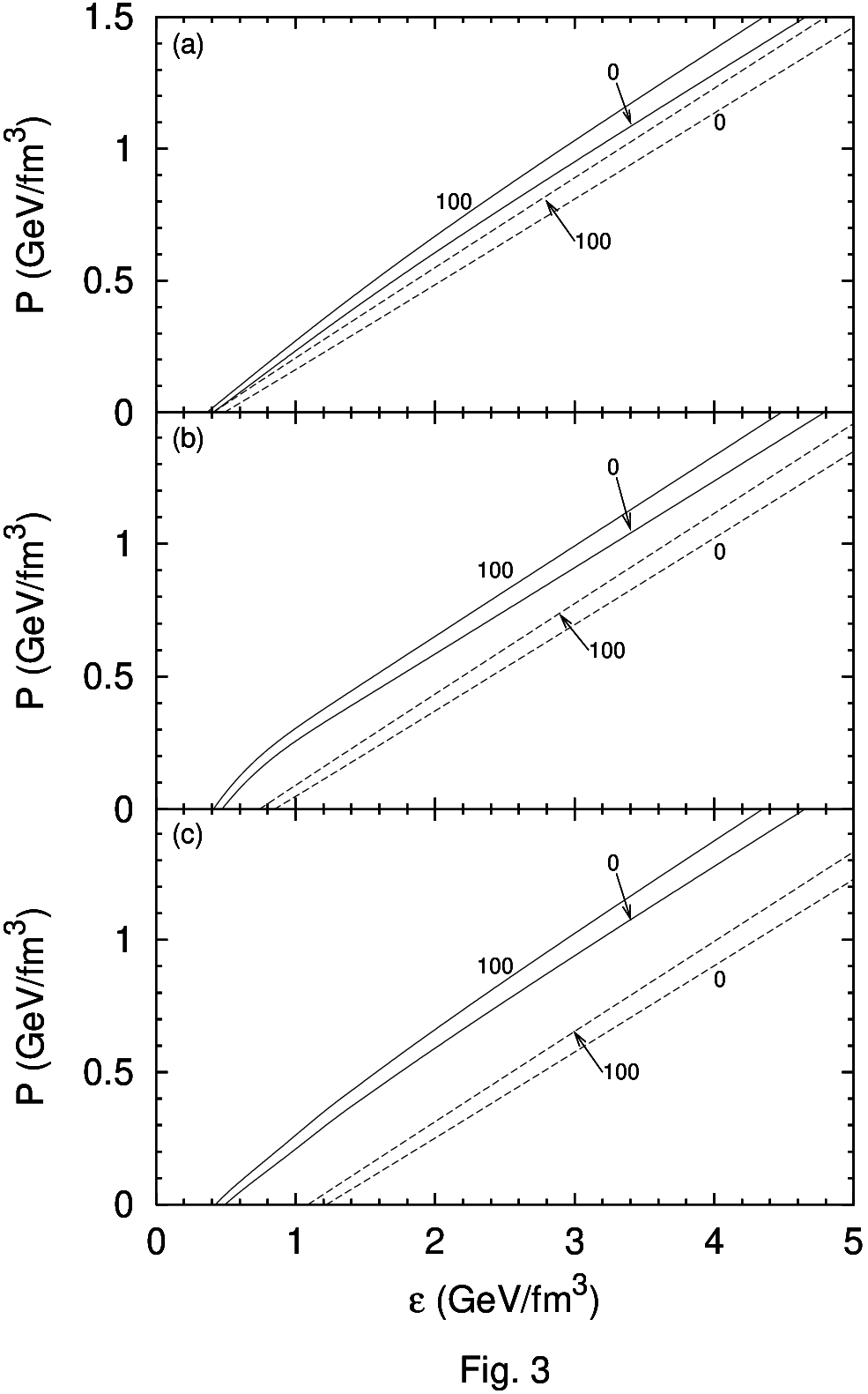}}}
%\caption{}
%\label{fig:...}
\end{figure}
\end{center}

%\newpage
\begin{center}
\begin{figure}[ht!]
%\centerline{\includegraphics*[width=\linewidth]{fig4.ps}}
\centerline{\resizebox{14cm}{!} {\includegraphics{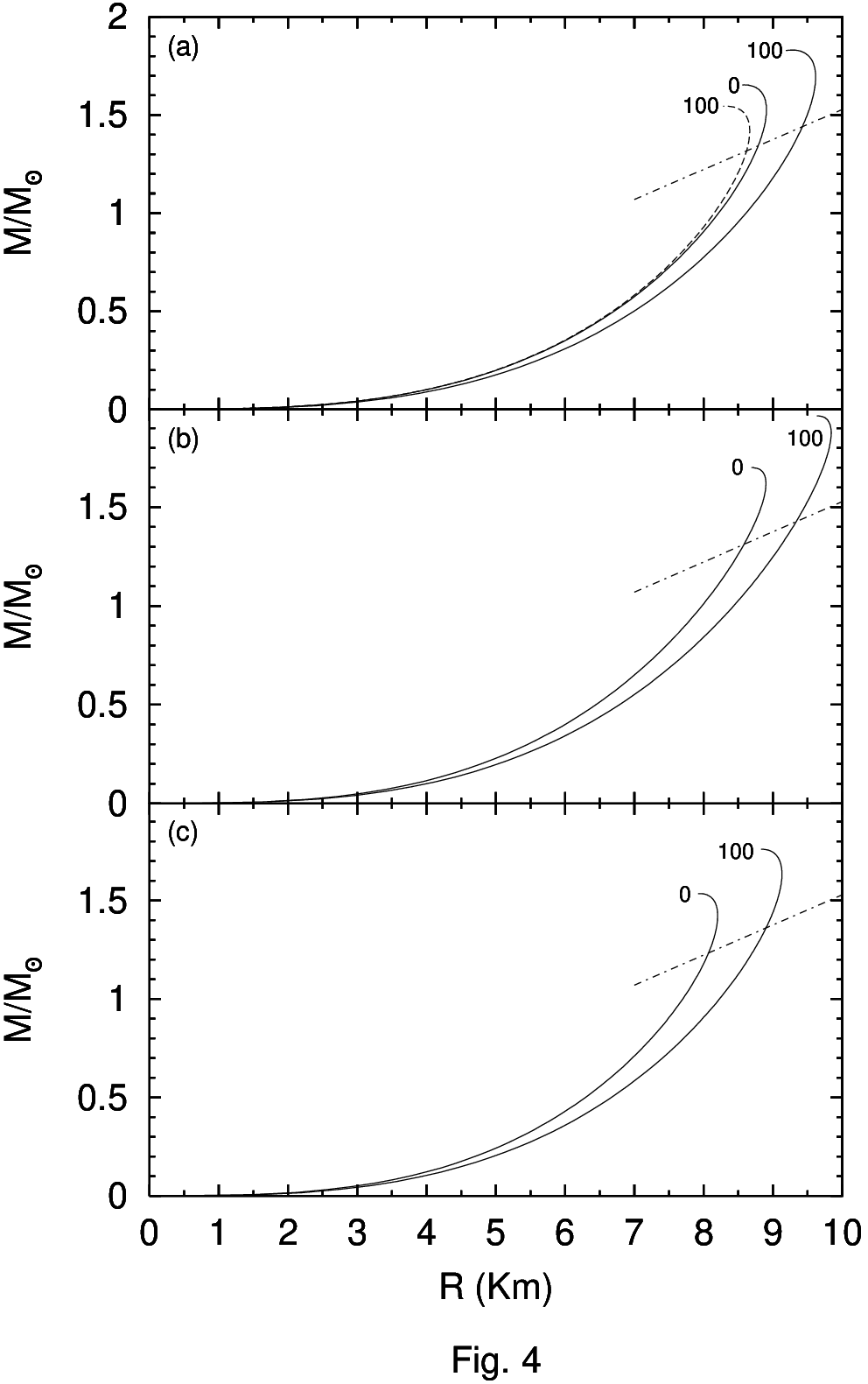}}}
%\caption{}
%\label{fig:...}
\end{figure}
\end{center}

%\newpage
\begin{center}
\begin{figure}[ht!]
%\centerline{\includegraphics*[width=\linewidth]{fig5.ps}}
\centerline{\resizebox{14cm}{!} {\includegraphics{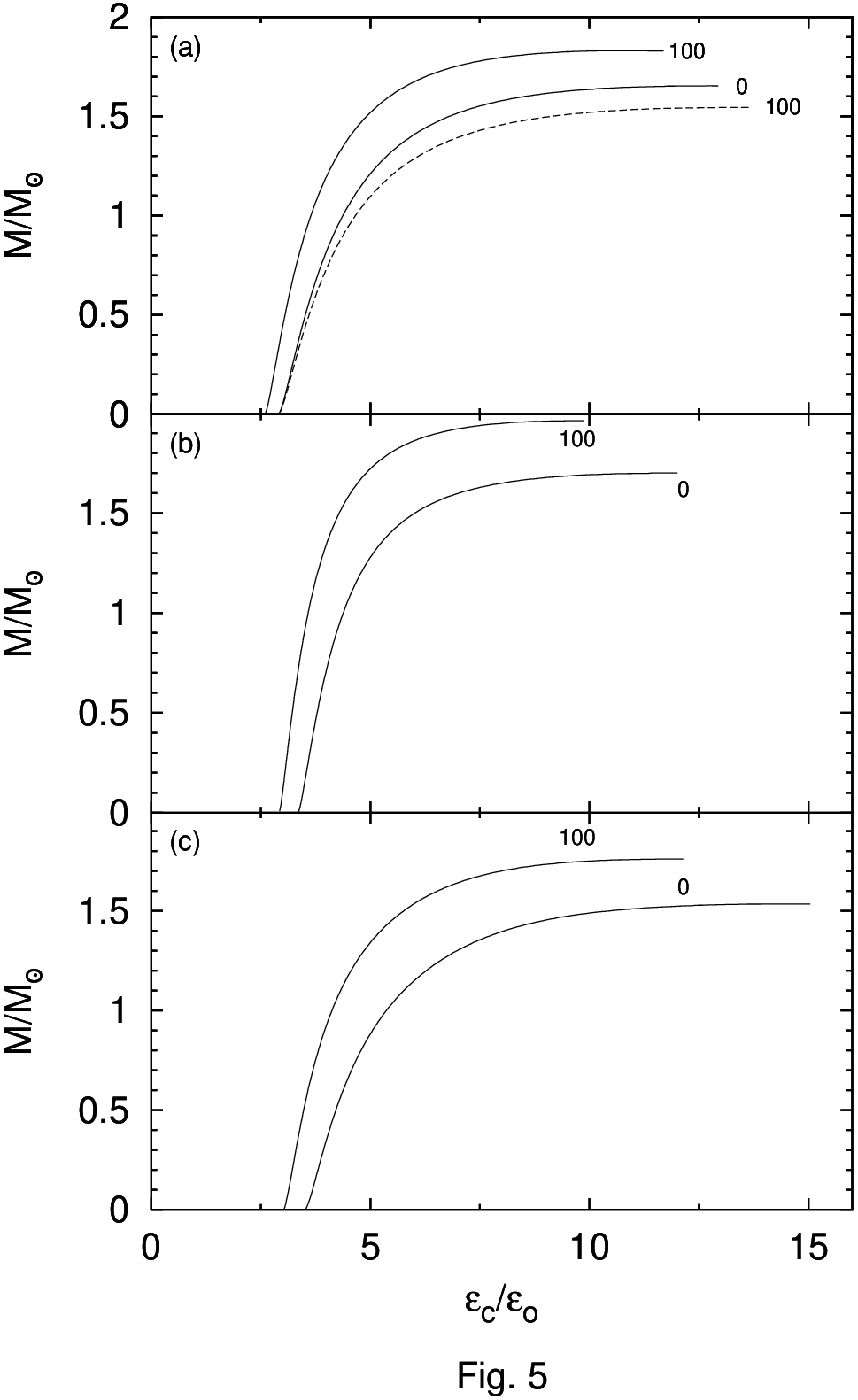}}}
%\caption{}
%\label{fig:...}
\end{figure}
\end{center}

%\newpage
\begin{center}
\begin{figure}[ht!]
%\centerline{\includegraphics*[width=\linewidth]{fig6.ps}}
\centerline{\resizebox{14cm}{!} {\includegraphics{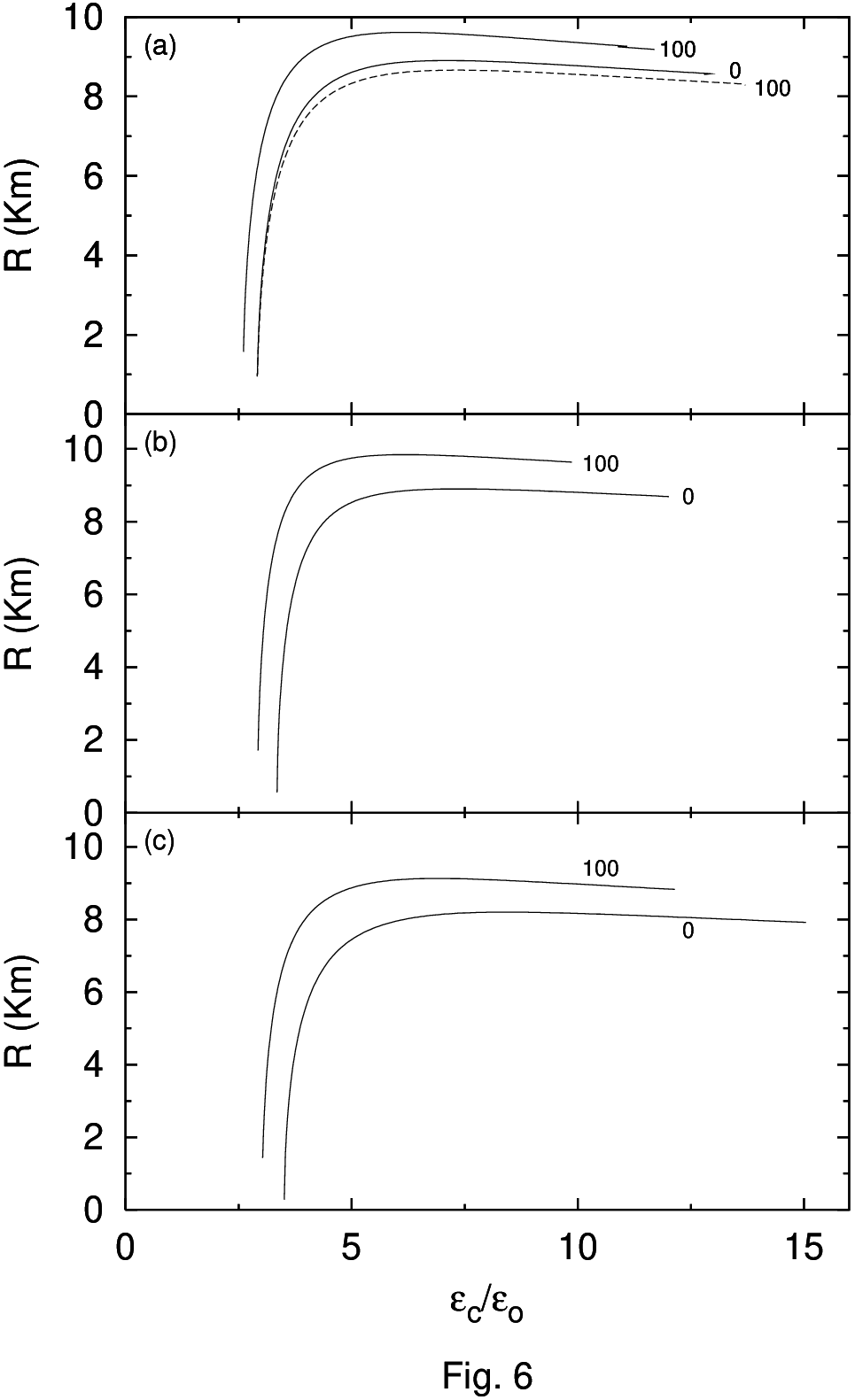}}}
%\caption{}
%\label{fig:...}
\end{figure}
\end{center}

%\newpage
\begin{center}
\begin{figure}[ht!]
%\centerline{\includegraphics*[width=\linewidth]{fig7.ps}}
\centerline{\resizebox{14cm}{!} {\includegraphics{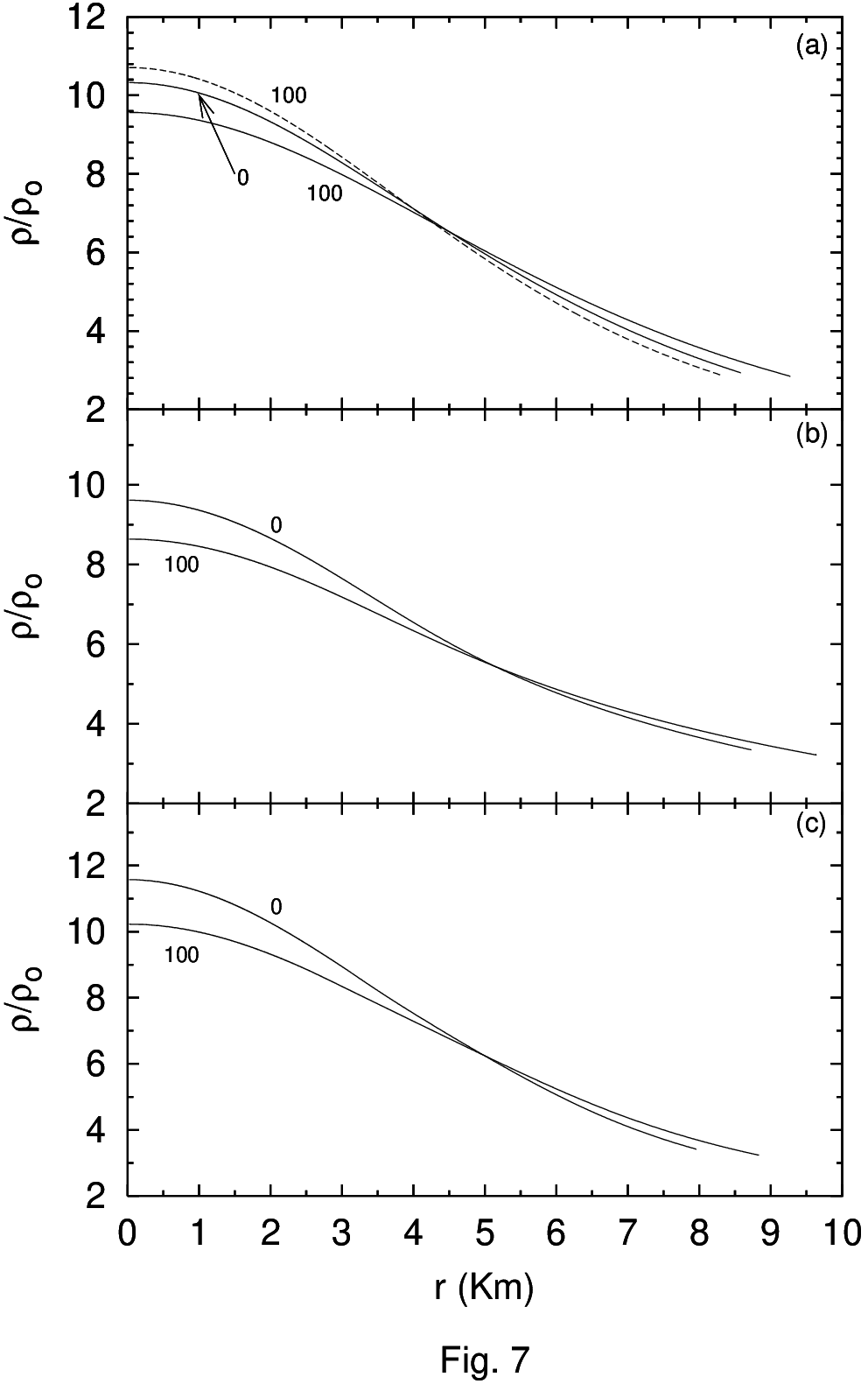}}}
%\caption{}
%\label{fig:...}
\end{figure}
\end{center}

%\newpage
\begin{center}
\begin{figure}[ht!]
%\centerline{\includegraphics*[width=\linewidth]{fig8.ps}}
\centerline{\resizebox{14cm}{!} {\includegraphics{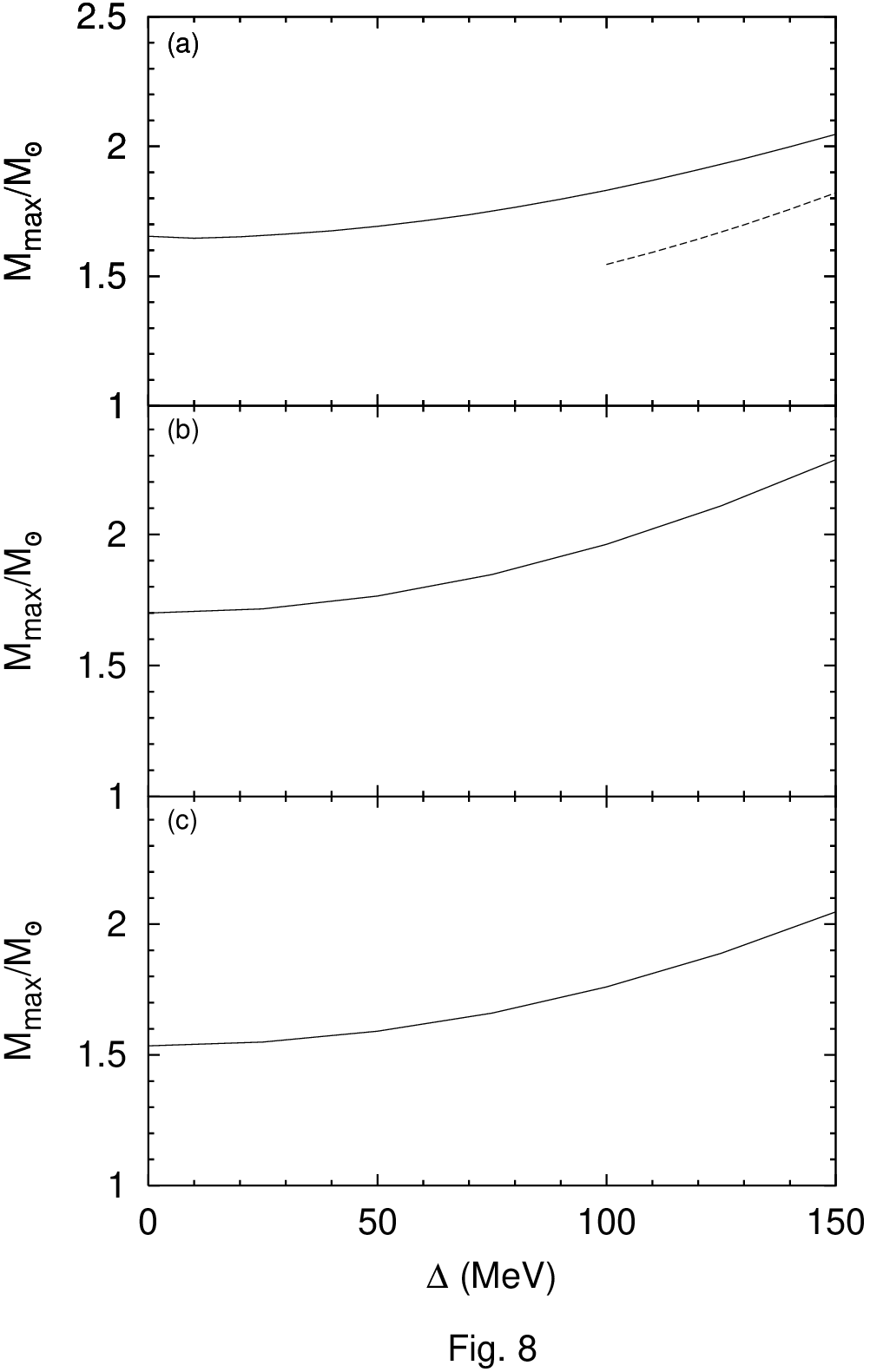}}}
%\caption{}
%\label{fig:...}
\end{figure}
\end{center}

%\newpage
\begin{center}
\begin{figure}[ht!]
%\centerline{\includegraphics*[width=\linewidth]{fig9.ps}}
\centerline{\resizebox{14cm}{!} {\includegraphics{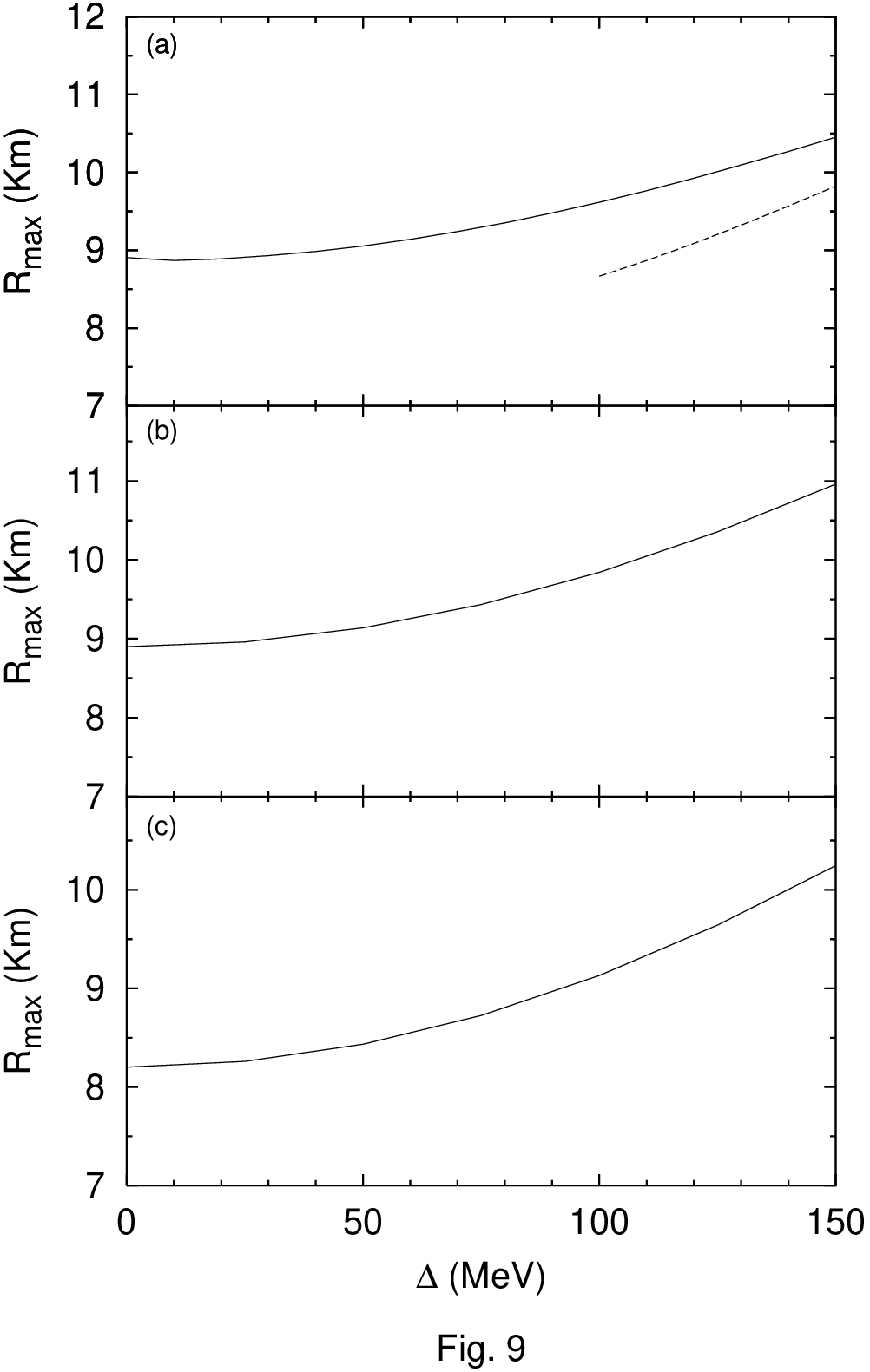}}}
%\caption{}
%\label{fig:...}
\end{figure}
\end{center}


\newpage

\begin{thebibliography}{99}

%\cite{Witten:1984rs}
\bibitem{Witten:1984rs}
E.~Witten,
%``Cosmic Separation Of Phases,''
Phys.\ Rev.\ D {\bf 30}, 272 (1984).
%%CITATION = PHRVA,D30,272;%%

%\cite{Bodmer:1971we}
\bibitem{Bodmer:1971we}
A.~R.~Bodmer,
%``Collapsed Nuclei,''
Phys.\ Rev.\ D {\bf 4}, 1601 (1971).
%%CITATION = PHRVA,D4,1601;%%

%\cite{Chin:1979yb}
\bibitem{Chin:1979yb}
S.~A.~Chin and A.~K.~Kerman,
%``Possible Longlived Hyperstrange Multi - Quark Droplets,''
Phys.\ Rev.\ Lett.\  {\bf 43}, 1292 (1979).
%%CITATION = PRLTA,43,1292;%%

%\cite{Farhi:1984qu}
\bibitem{Farhi:1984qu}
E.~Farhi and R.~L.~Jaffe,
%``Strange Matter,''
Phys.\ Rev.\ D {\bf 30}, 2379 (1984).
%%CITATION = PHRVA,D30,2379;%%

%\cite{Chodos:1974je}
\bibitem{Chodos:1974je}
A.~Chodos, R.~L.~Jaffe, K.~Johnson, C.~B.~Thorn and V.~F.~Weisskopf,
%``A New Extended Model Of Hadrons,''
Phys.\ Rev.\ D {\bf 9}, 3471 (1974).
%%CITATION = PHRVA,D9,3471;%%

%\cite{Haensel:qb}
\bibitem{Haensel:qb}
P.~Haensel, J.~L.~Zdunik and R.~Schaeffer,
%``Strange Quark Stars,''
Astron.\ Astrophys.\  {\bf 160}, 121 (1986).
%%CITATION = AAEJA,160,121;%%

%\cite{Alcock:1986hz}
\bibitem{Alcock:1986hz}
C.~Alcock, E.~Farhi and A.~Olinto,
%``Strange Stars,''
Astrophys.\ J.\  {\bf 310}, 261 (1986).
%%CITATION = ASJOA,310,261;%%

%\cite{Alford:1997zt}
\bibitem{Alford:1997zt}
M.~G.~Alford, K.~Rajagopal and F.~Wilczek,
%``QCD at finite baryon density: Nucleon droplets and color  
%superconductivity,''
Phys.\ Lett.\ B {\bf 422}, 247 (1998) [arXiv:hep-ph/9711395].
%%CITATION = HEP-PH 9711395;%%

%\cite{Rapp:1997zu}
\bibitem{Rapp:1997zu}
R.~Rapp, T.~Schafer, E.~V.~Shuryak and M.~Velkovsky,
%``Diquark Bose condensates in high density matter and instantons,''
Phys.\ Rev.\ Lett.\  {\bf 81}, 53 (1998)
[arXiv:hep-ph/9711396].
%%CITATION = HEP-PH 9711396;%%

%\cite{Rajagopal:2000wf}
\bibitem{Rajagopal:2000wf}
K.~Rajagopal and F.~Wilczek,
%``The condensed matter physics of QCD,''
arXiv:hep-ph/0011333.
%%CITATION = HEP-PH 0011333;%%

%\cite{Alford:1999pa}
\bibitem{Alford:1999pa}
M.~G.~Alford, J.~Berges and K.~Rajagopal,
%``Unlocking color and flavor in superconducting strange quark matter,''
Nucl.\ Phys.\ B {\bf 558}, 219 (1999)
[arXiv:hep-ph/9903502].
%%CITATION = HEP-PH 9903502;%%

%\cite{Schafer:1999pb}
\bibitem{Schafer:1999pb}
T.~Schafer and F.~Wilczek,
%``Quark description of hadronic phases,''
Phys.\ Rev.\ D {\bf 60}, 074014 (1999)
[arXiv:hep-ph/9903503].
%%CITATION = HEP-PH 9903503;%%

%\cite{Rajagopal:2000ff}
\bibitem{Rajagopal:2000ff}
K.~Rajagopal and F.~Wilczek,
%``Enforced electrical neutrality of the color-flavor locked phase,''
Phys.\ Rev.\ Lett.\  {\bf 86}, 3492 (2001)
[arXiv:hep-ph/0012039].
%%CITATION = HEP-PH 0012039;%%

%\cite{Steiner:2002gx}
\bibitem{Steiner:2002gx}
A.~W.~Steiner, S.~Reddy and M.~Prakash,
%``Color-neutral superconducting quark matter,''
Phys.\ Rev.\ D {\bf 66}, 094007 (2002)
[arXiv:hep-ph/0205201].
%%CITATION = HEP-PH 0205201;%%

%\cite{Alford:2002kj}
\bibitem{Alford:2002kj}
M.~Alford and K.~Rajagopal,
%``Absence of two-flavor color superconductivity in compact stars,''
JHEP {\bf 0206}, 031 (2002)
[arXiv:hep-ph/0204001].
%%CITATION = HEP-PH 0204001;%%

%\cite{Pons:2001px}
\bibitem{Pons:2001px}
J.~A.~Pons, F.~M.~Walter, J.~M.~Lattimer, M.~Prakash, R.~Neuhauser 
and P.~h.~An,
%``Towards a Mass and Radius Determination of the Nearby Isolated 
%Neutron Star RX J185635-3754,''
Astrophys.\ J.\  {\bf 564}, 981 (2002)
[arXiv:astro-ph/0107404].
%%CITATION = ASTRO-PH 0107404;%%

%\cite{Drake:2002bj}
\bibitem{Drake:2002bj}
J.~J.~Drake {\it et al.},
%``Is RXJ1856.5-3754 a quark star?,''
Astrophys.\ J.\  {\bf 572}, 996 (2002)
[arXiv:astro-ph/0204159].
%%CITATION = ASTRO-PH 0204159;%%

%\cite{Walter:2002uq}
\bibitem{Walter:2002uq}
F.~M.~Walter and J.~Lattimer,
%``A Revised Parallax and its Implications for RX J185635-3754,''
Astrophys.\ J.\  {\bf 576}, L145 (2002)
[arXiv:astro-ph/0204199].
%%CITATION = ASTRO-PH 0204199;%%

%\cite{Lugones:2002va}
\bibitem{Lugones:2002va}
G.~Lugones and J.~E.~Horvath,
%``Color-flavor locked strange matter,''
Phys.\ Rev.\ D {\bf 66}, 074017 (2002)
[arXiv:hep-ph/0211070].
%%CITATION = HEP-PH 0211070;%%

%\cite{Lugones:2002zd}
\bibitem{Lugones:2002zd}
G.~Lugones and J.~E.~Horvath,
%``High-density QCD pairing in compact star structure,''
Astron.\ Astrophys.\  {\bf 403}, 173 (2003).
[arXiv:astro-ph/0211638].
%%CITATION = ASTRO-PH 0211638;%%

%\cite{Muller:1980kf}
\bibitem{Muller:1980kf}
B.~Muller and J.~Rafelski,
%``Temperature Dependence Of The Bag Constant And The Effective 
% Lagrangian For Gauge Fields At Finite Temperatures,''
Phys.\ Lett.\ B {\bf 101}, 111 (1981).
%%CITATION = PHLTA,B101,111;%%

%\cite{Li:es}
\bibitem{Li:es}
S.~Li, R.~S.~Bhalerao and R.~K.~Bhaduri,
%``The Condensation Energy Of The N-J-L Vacuum And The Mit 
%Bag Constant,''
Int.\ J.\ Mod.\ Phys.\ A {\bf 6}, 501 (1991).
%%CITATION = IMPAE,A6,501;%%

%\cite{Reinhardt:tv}
\bibitem{Reinhardt:tv}
H.~Reinhardt and B.~V.~Dang,
%``Dynamical Bag Constant And Surface Tension Of Baryon Droplets 
% At Finite Temperature And Density,''
Phys.\ Lett.\ B {\bf 173}, 473 (1986).
%%CITATION = PHLTA,B173,473;%%

%\cite{Liu:2001em}
\bibitem{Liu:2001em}
Y.-x.~Liu, D.-f.~Gao and H.~Guo,
%``Density dependence of nucleon bag constant, radius and mass in 
%an  effective field theory model of QCD,''
Nucl.\ Phys.\ A {\bf 695}, 353 (2001)
[arXiv:hep-ph/0105202].
%%CITATION = HEP-PH 0105202;%%

%\cite{Roberts:1987xc}
\bibitem{Roberts:1987xc}
C.~D.~Roberts, R.~T.~Cahill and J.~Praschifka,
%``The Effective Action For The Goldstone Modes In A Global 
%Color Symmetry Model Of QCD,''
Ann. Phys. (NY)\ {\bf 188}, 20 (1988).
%%CITATION = APNYA,188,20;%%

%\cite{Burgio:2001mk}
\bibitem{Burgio:2001mk}
G.~F.~Burgio, M.~Baldo, P.~K.~Sahu, A.~B.~Santra and H.~J.~Schulze,
%``Maximum mass of neutron stars with a quark core,''
Phys.\ Lett.\ B {\bf 526}, 19 (2002)
[arXiv:astro-ph/0111440].
%%CITATION = ASTRO-PH 0111440;%%

%\cite{Aguirre:2002ws}
\bibitem{Aguirre:2002ws}
R.~Aguirre,
%``Chiral symmetry and strangeness content in nuclear physics 
%parametrized  by a medium dependent bag constant,''
Phys.\ Lett.\ B {\bf 559}, 207 (2003)
[arXiv:nucl-th/0212020].
%%CITATION = NUCL-TH 0212020;%%

%\cite{Aguirre:2003pc}
\bibitem{Aguirre:2003pc}
R.~Aguirre, private communication.

%\cite{Alford:2002rj}
\bibitem{Alford:2002rj}
M.~Alford and S.~Reddy,
%``Compact stars with color superconducting quark matter,''
Phys.\ Rev.\ D {\bf 67}, 074024 (2003)
[arXiv:nucl-th/0211046].
%%CITATION = NUCL-TH 0211046;%%

%\cite{Alford:2001zr}
\bibitem{Alford:2001zr}
M.~G.~Alford, K.~Rajagopal, S.~Reddy and F.~Wilczek,
%``The minimal CFL-nuclear interface,''
Phys.\ Rev.\ D {\bf 64}, 074017 (2001)
[arXiv:hep-ph/0105009].
%%CITATION = HEP-PH 0105009;%%

\bibitem{gle} N.K. Glendenning, {\it Compact Stars -- Nuclear
Physics, Particle Physics, and General Relativity}, 2nd edn.,
(Springer-Verlag, Berlin, 2000).

\bibitem{Tolman}
R.~Tolman, Phys.\ Rev.\ {\bf 55}, 364 (1939).

%\cite{Oppenheimer:1939ne}
\bibitem{Oppenheimer:1939ne}
J.~R.~Oppenheimer and G.~M.~Volkoff,
%``On Massive Neutron Cores,''
Phys.\ Rev.\  {\bf 55}, 374 (1939).
%%CITATION = PHRVA,55,374;%%

%\cite{Cottam:2002cu}
\bibitem{Cottam:2002cu}
J.~Cottam, F.~Paerels and M.~Mendez,
%``Gravitationally redshifted absorption lines in the X-ray burst 
%spectra of a neutron star,''
Nature {\bf 420}, 51 (2002)
[arXiv:astro-ph/0211126].
%%CITATION = ASTRO-PH 0211126;%%

\end{thebibliography}
\end{document}